\documentclass[aip,apl,reprint,nofootinbib]{revtex4-1}
\usepackage{amsmath}
\usepackage{amsfonts}
\usepackage{amssymb}
\usepackage{graphics}
\usepackage{float}
\usepackage{graphicx}
\usepackage{epstopdf}
\usepackage[compact]{titlesec}
\usepackage{natbib}
\usepackage{threeparttable}
\usepackage[titletoc,title]{appendix}
\usepackage{lipsum}
\usepackage{varwidth}
\usepackage{tabularx}

\begin{document}
\title{Spin-orbit coupling mediated tunable electron heat capacity of quantum wells}
\author{Parijat Sengupta}
\author{Enrico Bellotti}
\affiliation{Dept. of Electrical Engineering and Material Science Division \\
Boston University, Boston, MA 02215. 
}

\begin{abstract}
The heat capacity of conduction electrons obtained from the Sommerfeld expansion is shown to be tunable via the Rashba and Dresselhaus spin-orbit coupling parameters. Using an AlInSb/InSb/AlInSb as a representative heterostructure with alterable well and asymmetric barrier regions, the heat capacity is found to be higher for the spin-down electrons and suffers a reduction for wider wells. A further lowering is obtained through the application of an uniaxial strain. Finally, we suggest a method to determine the spin lifetimes for spins relaxing via the D'yakonov-Perel' mechanism from experimental estimates of thermodynamic potentials such as the Helmholtz free energy and the heat capacity.
\end{abstract}
\maketitle

The electronic contribution to the specific heat is usually masked by the phonon component or the lattice specific heat and becomes the dominant term only in the low-temperature regime. The specific heat measured much below the Debye and Fermi temperatures follows a relationship of the form $ \gamma T + \beta T^{3} $, the $ T^{3} $ contribution arising from lattice vibrations while the electrons contribute linearly in $ T $.~\cite{grosso2014solid} The electronic contribution is further reduced as few electron states are available in the thermal interval of the order $ k_{B}T $ around the Fermi energy. However, at sufficiently low temperatures, the measurement of the electronic specific heat unveils details about elementary excitations including features of the density of states close to the Fermi level, furnishing quantitative insight to several key microscopic processes. The many-body interactions in a strongly correlated Fermi liquid, for instance, is characterized by the deviation of Wilson's ratio (the ratio of the magnetic susceptibility to specific heat divided by temperature) from unity for weakly correlated electrons.~\cite{dressel2002electrodynamics} From a broader perspective, the laws of thermodynamics and statistical mechanics have been applied to explain the behaviour of a diverse body of phenomena such as Luttinger liquids, superfluidity, and superconductivity. Experimental groups routinely record data on the specific heat of superconductors to probe the superconducting energy gap, the order parameter symmetry, and the quasi-particle density of states.~\cite{wang2001specific} 

In this letter, in contrast to experimental measurements, we perform a theoretical determination of the electron heat capacity at constant volume, $ C_{v} $, of conduction electrons confined in the well region of zinc blende heterostructures grown along  $ \left[001\right] $-axis and show their tunability under extrinsic spin-orbit interactions. The tunability and optimization of heat capacity is fundamental to the design and the dynamic response of sensor materials, for e.g., thermal CMOS microtransducers, thermocouple thermometers, and low temperature sensors for ultra-scaled integrated circuits. We show that the heat capacity is affected by the heterostructure set-up whereby a wider well region reduces the $ C_{v} $ with the possibility of additional tuning through uniaxial strain which further lowers it. As an adjunct, we suggest a possible experimental method that facilitates the retrieval of spin lifetimes from experimentally recorded electron heat capacity and allied thermodynamic potentials.  

For low temperatures that comply with the criterion, $ k_{B}T \ll T_{F} $, where $ k_{B} = 8.617 \times 10^{-5}\,eV\,K^{-1} $ is the Boltzmann constant and $ T_{F} $ represents the Fermi temperature, the $ C_{v} $ can be reasonably approximated using the Sommerfeld expansion.~\cite{grosso2014solid} We begin by examining this quantity for conduction state electrons described by a parabolic dispersion and spin-split by the Rashba and Dresselhaus spin-orbit interaction (SOI).~\cite{winkler2003spin} The Rashba SOI manifested as splitting of energy bands typically originates in non-centrosymmetric crystals or structures with miscut surfaces, collectively referred to as structural inversion asymmetry (SIA). It is a remarkably adjustable phenomenon vital to the control and production of spin-polarized currents in spin-based devices. The Dresselhaus splitting on the other hand is intrinsic to the crystal and is generally an invariant parameter typically much smaller than Rashba SOI. Since these SOI mechanisms reorganize the electron energies, a concomitant effect on the $ C_{v} $ must exist. To see this, observe that a direct application of the Sommerfeld expansion leads to an SOI-governed density-of-states (DOS) dependent expression for the internal energy density $ \left(U\right)$ from which we extract the heat capacity using $ C_{v} = \partial\,U/\partial\,T $. It follows that the $ C_{v} $ is spin-dependent. Note that heat capacity always refers to a per unit area quantity.

\begin{figure}
\includegraphics[scale=0.6]{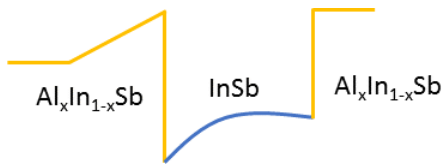}
\caption{Schematic of the asymmetric InSb quantum well surrounded by Al$_{x}$In$_{1-x}$Sb barriers of variable width and alloy composition. The dissimilar barriers flanking the quantum well induces a stronger SIA and enhanced Rashba splitting.}
\label{schema}
\vspace{-1.0em}
\end{figure}

We begin with an ensemble of conduction electrons in ZB quantum wells that are modeled assuming a parabolic dispersion with spin-orbit interaction terms. For ZB, the Hamiltonian has the form:
\begin{equation}
H_{zb} = \dfrac{p^{2}}{2m^{*}} + \lambda_{R}\left(\sigma_{x}k_{y} - \sigma_{y}k_{x}\right) + \lambda_{D}<k_{z}^{2}>\left(\sigma_{x}k_{x} - \sigma_{y}k_{y}\right),
\label{hzb}
\end{equation}
where $ \lambda_{R} > 0 $ and $ \lambda_{D} > 0 $ are the Rashba and Dresselhaus coupling parameters, respectively and the condition $ \lambda_{R} \neq \lambda_{D} $ holds for all cases. The cubic dependencies of the form $ k_{x}k_{y}^{2}\sigma_{x} - k_{y}k_{x}^{2}\sigma_{y} $ in the Dresselhaus Hamiltonian have been ignored while the $ k_{z} $ and $ k_{z}^{2} $ terms have been replaced by their quantized values 0 and $ \left(\pi/L\right)^{2} $. The width of the quantum well is $ L $ in appropriate unit. The dispersion relationship for the chiral bands of ZB conduction electrons using Eq.~\ref{hzb} is $ \hbar^{2}k^{2}/2m \pm \beta\,k $. The coefficient $ \beta $ takes the form $ \beta= \sqrt{\left(\lambda_{R}^{2} + \lambda_{D}^{2} + 2\lambda_{R}\,\lambda_{D}\,sin\,2\theta\right)} $ and $ \theta = tan^{-1}k_{y}/k_{x} $. Note that the effective mass in the above expressions is $ m^{*} $ which we obtain for several cases discussed later from relevant k.p Hamiltonians. For brevity, we will represent the eigen energies as $ \varepsilon\left(k\right) = \alpha k^{2} \pm \beta k $ where $ \alpha = \hbar^{2}/2m $. The eigen energies are chosen such that $ \varepsilon > 0 $. For numerical calculations, our representative ZB quantum wells are InSb quantum wells (see Fig.~\ref{schema}) surrounded by asymmetric Al$_{x}$In$_{1-x}$Sb barriers. The reason to select InSb as the well material lies in its narrow band gap, small effective mass, large \textit{g}-factor, and a significantly large spin-orbit coupling (SOC).

\begin{figure}[!t]
\includegraphics[scale=0.6]{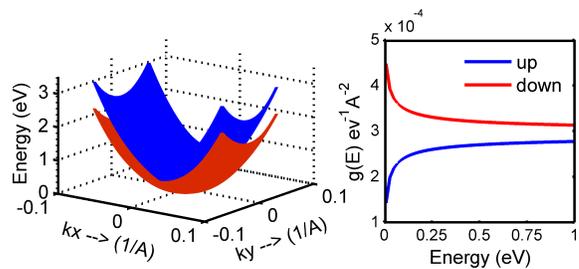}
\caption{The eigen energies using Eq.~\ref{hzb} is shown as a surface plot (left panel) along with the corresponding DOS for the two spin-split states. Note that at large values of energy when the $ ak^{2} $ term in the Hamiltonian is dominant, the DOS assumes a constant number in accord with the result for a 2D system. In calculating the eigen energies  and DOS, effective mass was assumed to be $ 0.014m_{0} $, where $ m_{0} $ is the free electron mass.}
\label{ekdos}
\vspace{-1.0em}
\end{figure}

To probe the tunability of the electron heat capacity, we first note that electron energy, $ U $, following the Sommerfeld expansion around the Fermi level $ \left(E = \mu\right)$ is
\begin{align}
U\left(T\right) = \int_{-\infty}^{\mu}ED\left(E\right)\,dE + \eta\dfrac{\partial}{\partial\,E}\left(ED\left(E\right)\right), 
\label{somm}       
\end{align}
where $ \eta = \left(\pi k_{B}T\right)^{2}/6 $. The DOS, $ D\left(E\right)$, is considered invariant to ambient conditions if  temperature-induced variations to the band structure which are usually insignificant are ignored. Notice that the Sommerfeld relation (Eq.~\ref{somm}) is not exact and we have chosen to ignore higher order temperature contributions.  From Eq.~\ref{somm}, the electronic contribution to the specific heat is $ C_{v} = \dfrac{\partial\,U}{\partial\,T} = \dfrac{\pi^{2}}{3}k_{B}^{2}TD\left(E\right) $. The DOS in Eq.~\ref{somm} can be calculated employing the standard result $ D\left(E\right) = \dfrac{1}{4\pi^{2}}\int\,d^{2}k\,\delta\left(E - \varepsilon\left(k\right)\right) $. Evaluating the integral gives $ D\left(E\right) = \dfrac{1}{4\pi^{2}}\int_{0}^{2\pi} d\theta\dfrac{k_{i}}{\vert\,-2\alpha k_{i} \mp \beta \vert} $, where $ k_{i} $ is a root of the equation $ g\left(k\right) = E - \alpha k^{2} \mp \beta k $. The upper (lower) sign is for the spin-up (down) branch. Here, two notable outcomes must be brought to attention: 1) The spin-down (-) branch has a larger DOS and therefore a higher population of electrons than the spin-up (+) branch. 2) The Fermi energy depends on the strength of the Rashba and Dresselhaus coupling parameters; doping apart, the spin-parameters can be used to reset the Fermi energy. As an illustration, through a numerical integration the DOS (and the connected dispersion) for the two spin split energy bands for an InSb quantum well of width $ L = 10.0\, nm $ is plotted in Fig.~\ref{ekdos}. To clearly show the spin splitting, the Rashba parameter was artificially set to $ 0.4 \,eV\,A $ and the Dresselhaus parameter was chosen to be $ \left( \pi/L\right)^{2} \times 0.48\, eVA^{3} $. Notice how with increasing energy, the DOS asymptotically approaches a constant value, which is the observed behaviour for a two-dimensional system. For quantitative expressions to supplement this discussion, see the section on Fermi energy, Ref.~\onlinecite{supl}.  

The heat capacity $ C_{v} $, using the Sommerfeld expansion, is therefore
\begin{equation}
C_{v} = \dfrac{k_{B}^{2}T}{12}\int_{0}^{2\pi}d\theta\dfrac{k_{i}}{\vert\,-2\alpha k_{i} \mp \beta \vert}.
\label{compcv}
\end{equation}
The upper (lower) sign gives the heat capacity for the ensemble of spin-up (down) electrons. The tunability of the heat capacity arises from the variable Rashba coupling coefficient and the conduction electron effective mass which enters the DOS expression. In particular, the strength of the Rashba coupling coefficient is $ \lambda = \lambda_{0}\langle\,E\left(z\right)\rangle $, where $ \langle\,E\left(z\right)\rangle $ serves as the average electric field. The material-dependent $ \lambda_{0} $ is given as~\cite{e1994spin}
\begin{equation}
\lambda_{0} = \dfrac{\hbar^{2}}{2m^{*}}\dfrac{\Delta}{E_{g}}\dfrac{2E_{g}+\Delta}{\left(E_{g} + \Delta\right)\left(3E_{g} + 2\Delta\right)}.
\label{rasz}
\end{equation}
In Eq.~\ref{rasz}, the band gap at $ \Gamma $ is $ E_{g} $, the spin-orbit splitting is $ \Delta $ and $ m^{*} $ is the effective mass at points in momentum space close to $ \Gamma $. The band gap and effective mass are obtained by diagonalization of Kane's eight-band Hamiltonian adapted for heterostructures.~\cite{bastard1990wave} The \textit{k.p} parameters in Ref.~\onlinecite{vurgaftman2001band} have been used. The effective mass and the band gap evidently change with confinement and external perturbations, such as mechanical strain, which in turn regulates the $ \alpha_{0} $ parameter. Further, notice that the Rashba coupling parameter is directly linked to the intrinsic spin-orbit coupling and is enhanced by a reduction in the effective band gap, $ E_{g} $. A pronounced SOC, for instance, in the layered polar semiconductor BiTeI~\cite{ishizaka2011giant,manchon2015new} or a strong local electric field, such as the one discovered in the cubic perovskite strontium titanate (SrTiO$_{3}$) give rise to large energy splittings of the order $ 100.0\, meV $.~\cite{santander2014giant} It is useful to bear in mind that while a finite Rashba SOC is usually found in asymmetric crystals, a localized asymmetry arising from strain or a strongly confined impurity can also induce a discernible Rashba splitting as reported for the centrosymmetric layered material 2H-WSe$_{2}$.~\cite{zhang2014hidden} Lastly, notice that the electric field acting via the gate adjusts the strength of the Rashba coefficient; for numerical calculations the electric field has been set to $ 10^{6}\,V/m $, a value achieved either through an external bias or doping. 

\begin{figure}
\includegraphics[scale=0.65]{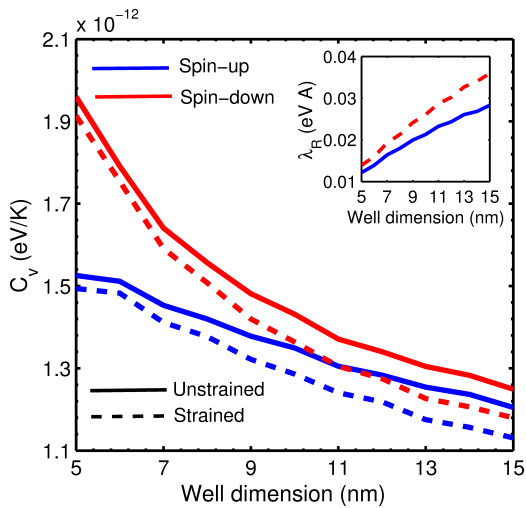}
\caption{The calculated heat capacity of conduction electrons for varying InSb well dimensions flanked by AlInSb alloy (see Fig.~\ref{schema}). The Fermi energy was set to $ \epsilon_{f} = 4.0\, meV $ and temperature is $ T = 5\, K $.  Note that the Fermi temperature $ T_{f}\left(\epsilon_{f}/k_{B}\right) = 46\,K  \gg T $ which renders the Sommerfeld expansion valid. The inset shows the Rashba parameter calculated using Eq.~\ref{rasz}.}
\label{rcv}
\vspace{-1.0em}
\end{figure}

An adjustable asymmetry therefore can alter the RSOC which in turn (this follows directly by an examination of Eqs.~\ref{compcv},~\ref{rasz}) manifests in a variable heat capacity. In our simulations, the asymmetry inducements are supplied through strain, dimensional confinement, and an irregular heterostructure configuration (see Fig.~\ref{schema}). The asymmetry through the heterostructure is introduced by employing variable widths and alloy composition $ \left(x\right) $ for the left and right barrier (Al$_{x}$In$_{1-x}$Sb) while a uniaxial strain further accentuates the underlying crystal order. The effective band gap in Eq.~\ref{rasz} is primarily modulated through multiple dimensional confinement widths. For purpose of calculation, the mole fraction for left- and right-barriers have been set to 0.25 and 0.45, respectively while the well widths are also dissimilar to magnify the local asymmetry. A stress of 1.5 GPA acts along the $ \left[001\right] $-axis. For detailed band structure calculations, see Ref.~\onlinecite{supl} and Ref.~\onlinecite{sengupta2016numerical}. With this in mind, let's now examine Fig.~\ref{rcv} which plots the electron heat capacity for heterostructures with well (InSb) regions of varying confinement. The alloyed barrier on the left (right) of the well is $ 5.0\,(7.0)\,nm $ and the molar composition $ \left( x\right) $ was set to $ 0.25\,(0.55) $. 
  
Firstly, we notice that the Rashba parameter, $ \lambda_{R} $, is sensitive (see inset, Fig.~\ref{rcv}) to dimensional confinement and strain; an increase in the well dimension and uniaxial strain enhances it. To explain this increase, recall that strain builds up asymmetry within the heterostructure augmenting $ \lambda_{R} $, while a wider well region has smaller band gap entailing a stronger SOI. A stronger SOI is a cause for enlarged Rashba splitting.~\cite{zawadzki2003spin} The boost to the Rashba splitting induced through strain and confinement, however, manifests as a change to the DOS; the DOS for the spin-up states show a decrease while their spin-down counterparts have a reverse behaviour (see Eq. 3 and a short discussion therein, Ref.~\onlinecite{supl}). A straightforward manifestation of this is the ascendancy of the heat capacity for the spin-down electrons over the oppositely spin-polarized set, since the former has a higher DOS, an entity that directly links to heat capacity through Eq.~\ref{compcv}. We also observe that the well dimension impacts the $ C_{v} $ which is borne out by the fact that the effective mass of conduction electrons and consequently the DOS is adjusted; a wider well has lower effective mass yielding a lower DOS and therefore a lower heat capacity. It is thus evident that a tuning of the heat capacity to a desired level can be achieved through an application of strain or a choice of heterostructure dimensions; both of which modify the DOS through changed effective mass and the Rashba parameter. A more dynamic method, however, is to alter the Rashba coupling strength $\left(\lambda_{R} = \lambda_{0}E\right) $ through an external gate bias.

The tunability of the electron heat capacity under spin-orbit interaction has been the main focus of this letter; however, we can also turn the argument and instead ask if it is feasible to retrieve any observable from a direct measurement of the heat capacity. This part is motivated by recent measurements~\cite{jk} of an exceptionally high heat capacity, reflected in a large effective mass, in compounds collectively called heavy fermion systems. In line with such measurements, we suggest that a similar observable, the spin lifetime, can be determined from experimentally available heat capacity data. For a spin-ensemble relaxing via the D'yakanov-Perel' (DPM) mechanism, the average spin equation is~\cite{vzutic2004spintronics}
\begin{equation}
\langle \dot{S_{i}}\rangle = -\tau_{p}\left[\langle \mathbf{\Omega}^{2}\rangle\langle S_{i}\rangle - \sum\limits_{j}\langle \Omega_{i}\Omega_{j}\rangle \langle S_{i}\rangle\right].
\label{speq}
\end{equation}
In Eq.~\ref{speq}, the spin precession frequency is  $ \mathbf{\Omega}\left(k\right) $ and the momentum relaxation time $ \tau_{p} $ is independent of energy. In general, spins relax via DPM in the diffusive limit when the momentum relaxation time is considerably lesser than for a complete spin rotation on the Bloch sphere. For a two-dimensional spin ensemble that precesses under the combined spin-orbit magnetic field of Rashba and Dresselhaus interaction, the $ \mathbf{\Omega}\left(k\right) $ dependence in each case is $ \mathbf{\Omega}_{R}\left(k\right) = \lambda_{R}\left(k_{y}, -k_{x}\right) $ and $ \mathbf{\Omega}_{D}\left(k\right) = \lambda_{D}\left(k_{x}, -k_{y}\right) $, respectively. The spin dephasing tensor by substituting the spin precession components in Eq.~\ref{speq} for an asymmetric quantum well grown along $ \left(001\right)-$axis is~\cite{averkiev2002spin}
\begin{equation}
\dfrac{1}{\tau_{z}} = \dfrac{\tau_{p}k^{2}}{\hbar^{2}}\left(\lambda_{R}^{2} + \lambda_{D}^{2}\right);\dfrac{1}{\tau_{\pm}} = \dfrac{\tau_{p}k^{2}}{2\hbar^{2}}\left(\lambda_{R} \pm \lambda_{D}\right)^{2}.
\label{splft}
\end{equation}
The subscripts $ \pm $ denote the $ \left(110\right) $ and $ \left(1\overline{1}0\right) $ orientation. An experimental procedure to measure spin lifetime must therefore entail the direct substitution of $ C_{v} $ data in Eq.~\ref{splft}. Using Eq.~\ref{compcv}, it is possible to write the Rashba coupling parameter in Eq.~\ref{splft} as $  \lambda_{R} \approx  \sqrt{\left(4\alpha \Lambda E\right)/\left(1-\Lambda\right)} $, where $ \Lambda = \left[1 - \left(12\alpha C_{v}\right)/\left(\pi k^{2}_{B}T\right)\right]^{2} $. A proof is added to the supplementary material, Ref.~\onlinecite{supl}.    

The evaluation of Eq.~\ref{splft} requires the momentum scattering time; the mometum scattering at low temperatures is primarily impurity scattering driven as the suppressed phonon modes contribute little to electron-phonon scattering. Taking this in to account, the imaginary part of the retarded self-energy $ \left(\Sigma\right) $ for the ensemble of conduction electrons whose motion is impeded by impurity sources, allows us to estimate the scattering time $\left(\tau_{p}\right) $ through the relation, $  1/\tau_{p} = \left(2/\hbar\right)Im\Sigma $. The retarded self-energy in the self consistent Born approximation (SCBA) is expressed as a pair of equations~\cite{di2015statistical}
\begin{equation}
\begin{aligned}
\label{scba1}
G_{ks}\left(\epsilon\right) = \dfrac{1}{\epsilon - \epsilon_{ks} - \Sigma\left(\epsilon\right)};
\Sigma\left(\epsilon\right) = n_{i}v_{i}^{2}\int\,\dfrac{d^{2}k}{4\pi^{2}}G_{ks}\left(\epsilon\right),
\end{aligned}
\end{equation}
where $ n_{i} $ and $ v_{i} $ denote the density and strength of impurities, respectively and $ G_{ks}\left(\epsilon\right) $ is the $ 2 \times 2 $ retarded Green's function diagonal with respect to the band index \textit{s} ($\langle\,s\vert\,G_{k}\left(\epsilon\right)\vert\,s\rangle = \delta_{ss^{'}}G_{ks}\left(\epsilon\right) $). The retarded self-energy, $ \Sigma $, in SCBA averaged over impurity distributions is also diagonal with respect to the band index \textit{s} and independent of \textbf{\textit{k}}. 
Using the Green's function, $ \left(E\mathbb{I} - H_{zb}\right)^{-1} $, corresponding to the Hamiltonian in Eq.~\ref{hzb}, the imaginary part of the retarded self-energy is
\begin{align}
Im\Sigma\left(E\right) &= \dfrac{n_{i}v_{i}^{2}}{8\pi}\int d^{2}k\biggl[\left (\delta_{1} + \delta_{2} \right) \pm  \left (\delta_{1} - \delta_{2} \right)\biggr],
\label{selfe2}
\end{align}
where $ \delta_{1}\left(\delta_{2}\right)=\delta\left(E - \alpha k^{2} \mp \beta k\right) $. In deriving Eq.~\ref{selfe2}, the standard relation $ \dfrac{1}{x \pm i\delta} =  \mathbb{P}\dfrac{1}{x} \mp i\pi\delta\left(x\right) $ was used and as usual, the upper (lower) sign is for the spin-up (down) band. Evaluating this integral which closely resembles the one we encountered while computing the DOS, the retarded self-energy is approximately given as $ n_{i}m^{*}v_{i}^{2}/4\hbar^{2} $ for both set of spin-chiral bands (see section on SCBA, Ref.~\onlinecite{supl}). For an experimentally preset impurity density, $ n_{i} = 1.0 \times 10^{11}\,cm^{-2} $ and the attendant impurity potential~\cite{sengupta2016photo} being $ 0.1\,keV \AA^{2} $, the momentum scattering time for conduction electrons with effective mass equal to 0.0201 (obtained from a \textit{k.p} calculation for a 10.0 $\mathrm{nm} $) wide well works out to roughly 5 ps. A sample calculation of spin lifetimes where the measured $ C_{v} $ is $ 5.2 \times 10^{-11}\, eV/K $ (this is extracted for purpose of illustration from data presented in Fig.~\ref{rcv}) gives $ \tau_{z} $ as 5.68\, ps while the lifetimes along the $ \left(110\right) $ and $ \left(1\overline{1}0\right) $ directions, $ \tau_{+} $ and $ \tau_{-} $, are 10.98 and 11.78 ps, respectively. A comment is in order here : Heat capacity is one of the several thermodynamic potentials that we have used in our analysis of spin lifetime. A similar estimation of the spin lifetime can be performed if we consider another thermodynamic potential, the Helmholtz free energy, defined as $ F = U\left(T = 0\right) - TS $, where $ S = \int_{0}^{T}C_{v}/T^{'}\,dT^{'}  $ is the entropy. At two distinct temperatures, the difference between the Helmholtz free energies is $ \Delta F = \left(T_{2} - T_{1}\right)\Delta S $, where $ \Delta S = 0.5\left(S_{1} + S_{2}\right) $. The entropy $ S_{i} $ at temperature $ T_{i} $, in the Sommerfeld expansion is identical to $ C_{v} $.~\cite{grosso2014solid}

To summarize, we have demonstrated the SOI-governed tunability of the electron heat capacity. The Rashba SOI, primarily, is adjusted utilizing dimensional confinement and mechanical strain. We have also shown that thermodynamic potential measurements offer an experimental technique to gauge the spin lifetimes. An application of our results that we did not discuss in the text is the computation of electronic thermal conductivity, $ \kappa_{e} = C_{v} v_{f}^{^{2}}\tau_{p}/3 $, where $ v_{f} $ is the Fermi velocity. Going a step further, following the Wiedemann-Franz law which relates the thermal conductivity to electrical conductivity $ \left( \sigma\right)$ at low temperatures through the empirical relation, $ \kappa/\sigma = L T $, where $ L $ is the Lorentz number, it is possible to evaluate the electrical resistivity $ \left(1/\sigma\right) $ from heat capacity measurements. Lastly, note that we have  assumed $ \lambda_{R} \neq \lambda_{D} $, however, the behaviour of the thermodynamic potentials under their equality hallmarked by the persistent spin helix~\cite{sengupta2016tunable,liu2006persistent} and a much longer spin lifetime has not been examined. 

%

\end{document}